\documentclass{mn2e}
\usepackage{epsf,times}
\newcommand{\etal}{{et al}\/.}
\begin{document}
\title[Hot and cold gas accretion]{Hot and cold gas accretion and feedback in radio-loud active galaxies}
\author[M.J.~Hardcastle \etal]{M.J.\ Hardcastle$^1$, D.A. Evans$^2$
  and J.H. Croston$^1$\\
$^1$ School of Physics,
  Astronomy and Mathematics, University of
Hertfordshire, College Lane, Hatfield, Hertfordshire AL10 9AB\\
$^2$ Harvard-Smithsonian Center for Astrophysics, 60 Garden Street, Cambridge, MA~02138, USA}
\maketitle
\begin{abstract}
We have recently shown that X-ray observations of the population of
`low-excitation' radio galaxies, which includes most low-power,
Fanaroff-Riley class I sources as well as some more powerful
Fanaroff-Riley class II objects, are consistent with a model in which
the active nuclei of these objects are not radiatively efficient at
any waveband. In another recent paper Allen et al. have shown that
Bondi accretion of the hot, X-ray emitting phase of the intergalactic
medium (IGM) is sufficient to power the jets of several nearby,
low-power radio galaxies at the centres of clusters. In this paper we
combine these ideas and suggest that accretion of the hot phase of the
IGM is sufficient to power {\it all} low-excitation radio sources,
while high-excitation sources are powered by accretion of cold gas
that is in general unrelated to the hot IGM. This model explains a
number of properties of the radio-loud active galaxy population, and
has important implications for the energy input of radio-loud active
galactic nuclei into the hot phase of the IGM: the energy supply of
powerful high-excitation sources does not have a direct connection to
the hot phase.
\end{abstract}


\section{Introduction}
In the conventional picture of active galactic nuclei (AGN), accretion
of cold matter on to the central supermassive black hole of a galaxy
proceeds by way of a luminous accretion disc: the disc provides the
radiation field that photoionizes the broad-line region (BLR) and
narrow-line region (NLR) in the optical and gives rise to the X-ray
emission via Compton scattering. Without radiatively efficient
accretion via the disc, none of these standard features of such an AGN
would be observed. Unified models propose that a direct view of
the BLR, the optical continuum, and the soft X-rays may be obscured
(e.g. in Seyfert 2s) by a dusty `torus': but in this case the torus
re-radiates strongly in the mid-IR band, while the hard X-rays are
still detectable, so that the presence of a luminous AGN can still be
inferred. Heavily obscured nuclear X-ray emission combined with mid-IR
radiation is strong evidence for an obscuring torus.

Although this simple picture is broadly consistent with observations
of radio-quiet AGN, it has been known for some time (Hine \& Longair
1979) that it is not sufficient to explain the properties of
radio-loud objects -- the radio galaxies and quasars. By analogy with
radio-quiet objects, we would expect that face-on radio-loud objects
(the broad-line radio galaxies and radio-loud quasars) would show
optical continuum emission, unabsorbed X-rays, and broad and narrow
optical lines, while edge-on radio-loud objects (narrow-line radio
galaxies, NLRG) would show narrow lines only in the optical, would
have heavily absorbed nuclear X-rays, and would have a clear
mid-infrared signature of the absorbing torus. This radio-loud unified
picture does indeed seem to describe the nuclei of many of the most
powerful radio sources (e.g. Barthel 1989; Haas \etal\ 2004), although
an additional jet-related X-ray component must be present to explain
the nuclear soft X-ray detections of the radio galaxies (e.g.
Hardcastle \& Worrall 1999; Belsole \etal\ 2006). However, Hine \&
Longair observed, in work more recently confirmed by others (e.g.
Laing \etal\ 1994; Jackson \& Rawlings 1997), that many radio galaxies
do not have the strong high-excitation narrow-line optical emission
that is expected from a conventional AGN. The objects lacking these
narrow lines, which we shall refer to here as low-excitation radio
galaxies (LERGs) are commonest at low radio luminosities: indeed,
almost all low radio power (Fanaroff-Riley class I, or FRI) radio
galaxies are LERGs. But in samples of radio galaxies the LERG
phenomenon persists to high radio luminosities; many powerful FRII
radio sources are LERGs as well. LERGs in general show no evidence in
the mid-IR for an obscuring torus, either at low or high luminosities
(Whysong \& Antonucci 2004; Ogle \etal\ 2006) and their optical nuclei
are consistent with being dominated purely by jet-related emission
(Chiaberge \etal\ 2002). The relation between their emission-line and
radio properties is also different (Baum, Zirbel \& O'Dea 1995). Most
recently, we have argued (Evans \etal\ 2006; Hardcastle \etal\ 2006,
hereafter H06) that both low-power FRI and high-power FRII LERGs show
no evidence for accretion-related X-ray emission, absorbed or
unabsorbed, over and above what is likely to originate in the nuclear
(pc-scale) jets. It seems most likely that the LERG population simply
does not have a radiatively efficient accretion flow, and so produces
none of the optical or X-ray characteristics of a conventional AGN. In
other words, we and others have argued that {\it LERGs may be a class
of luminous active galaxies that accrete radiatively inefficiently,
with almost all the available energy from accretion being channelled
into the jets}.

In this paper we take this argument a step further. We show that it
is possible that these apparently different accretion modes may be a
result of a different {\it source} for the accreting gas, building on
the recent result of Allen \etal\ (2006), who showed that some
low-luminosity radio galaxies in the centres of clusters could be
powered by Bondi accretion from the hot, X-ray emitting medium, and
supporting arguments about the nature of the accretion mode in
low-power radio sources recently made by Best \etal\ (2006). We
propose that the low-excitation objects are fuelled by accretion from
the hot phase, while the high-excitation objects require fuelling from
cold gas. If different types of radio sources do have different
accretion modes, then a number of features of the radio-loud AGN
population can be explained; there are also important implications for
the so-called `feedback' process in which radio-loud AGN do work on
their hot-gas environments. We begin (Section \ref{bondi}) by using
some simple quantitative tests to show that it is at least plausible
that accretion from the hot phase can fuel even powerful
low-excitation radio sources, but that it is hard to see how it can
power the most powerful FRIIs. In Section \ref{implications} we
discuss some of the wider implications of this model. We summarize our
results and discuss future tests of the model in Section
\ref{summary}.

Throughout the paper we use a cosmology with $H_0 = 70$ km s$^{-1}$
Mpc$^{-1}$, $\Omega_{\rm m} = 0.3$ and $\Omega_\Lambda = 0.7$.

\section{Bondi accretion and powerful radio sources}
\label{bondi}

Without committing ourselves to a particular model of the way in which
the accreted material actually powers the jets, we may use the Bondi
accretion rate to estimate the amount of power that the central
supermassive black hole can extract from accretion of the hot phase of
the IGM, following the analysis of Allen \etal\ (2006). The Bondi rate is given by
\begin{equation}
\dot M = \pi\lambda c_{\rm s} \rho_{\rm A} r_{\rm A}^2
\end{equation}
where $\lambda$ is a constant which has the value 0.25 for an
adiabatic index $5/3$, $c_{\rm s}$ is the sound speed in the
medium ($c_{\rm s} = \sqrt{\Gamma kT/\mu m_{\rm p}}$), $r_{\rm A}$ is
the Bondi accretion radius and $\rho_{\rm A}$ is the
density at that radius. $r_{\rm A}$ is given by
\begin{equation}
r_{\rm A} = {{2GM_{\rm BH}}\over{c_s^2}}
\label{rbondi}
\end{equation}
so that, for $\lambda=0.25$, we have the simple form
\begin{equation}
\dot M = \pi\rho_{\rm A} G^2 M_{\rm BH}^2 / c_s^3
\label{mdot}
\end{equation}
and we assume, again following Allen \etal , that all material that
accretes across the Bondi radius is accreted on to the black hole, so
that the available power for AGN activity, $P_{\rm B} = \eta \dot M
c^2$, where $\eta$ is an efficiency factor. To show that the jet can
be powered by accretion of the IGM, we require that the jet power $Q
\la P_{\rm B}$.

Allen \etal\ (2006) were able to show that Bondi accretion could
supply enough energy to power the radio sources they studied because
they were able to estimate the power in the jets $Q$ from the
timescales and energies required to inflate observed cavities in the
external medium, and the density at the Bondi radius $\rho_{\rm
A}$ and the temperatures $kT$ by extrapolation from observations. They
obtained black hole masses from the mass-velocity dispersion relation
($M_{\rm BH}$ - $\sigma$ relation) of Tremaine \etal\ (2002), except
in the case of M87 for which a direct $M_{\rm BH}$ measurement was
available. For powerful, distant radio galaxies in poorer
environments, we generally do not have either a direct measurement of
$Q$, an $M_{\rm BH}$ measurement, a galactic velocity dispersion
$\sigma$, or a particularly good estimate of the density at the
Bondi radius $\rho_{\rm A}$, and so more indirect methods must be
used.

It is well known that radio observations alone cannot be used to
measure $Q$ for a given radio galaxy without making numerous
assumptions about the particle and field content of the lobes, the
validity of conventional spectral ageing techniques and the fraction
of the jet power that has gone into doing work on the external medium.
However, there is a class of low-power objects where a model-dependent but
robust jet power measurement can be made. These are the twin-jet FRI
sources, where the behaviour of surface brightness and polarization in
the (presumed intrinsically symmetrical) jet and counterjet can be
used to model jet dynamics (Laing \& Bridle 2002a; Canvin \etal\ 2005;
Laing \etal\ 2006). When combined with X-ray information on the
pressure gradient surrounding the jets the
dynamical model allows a jet power to be derived. The only source to
have a published jet power derived from this method is 3C\,31 (Laing
\& Bridle 2002b) but since 3C\,31 is one of the archetypes of its
class, and has a relatively powerful jet for an FRI, we begin by
considering it in more detail.

To derive black-hole masses for radio galaxies we use the relationship
between $M_{\rm BH}$ and K-band absolute bulge magnitude derived by
Marconi \& Hunt (2003) for nearby sources:
\begin{equation}
\log_{10} M_{\rm BH} = 8.21 + 1.13 \times (\log_{10} L_{\rm K} - 10.9)
\end{equation}
where $L_{\rm K}$ is the K-band luminosity in solar units. We choose
to use this particular relation because it is well calibrated against
an $M_{\rm BH}$ - $\sigma$ relation -- indeed, Marconi \& Hunt argue
that the dispersion in the $M_{\rm BH}$ - $K$ relation is as small as
that for the $M_{\rm BH}$ - $\sigma$ relation for objects with
well-determined black hole masses -- and because K-band magnitudes for
powerful radio sources are readily available. Marconi \& Hunt carried
out a bulge-disc decomposition for their objects, which included a
number of spirals, but we assume that the total K-band magnitude is
the appropriate number to use for the elliptical galaxies that host
radio sources. The observed scatter in the $M_{\rm BH}$ - $K$ relation
is about 0.5 dex, implying (eq.\ \ref{mdot}) an uncertainty of about
one order of magnitude in the available Bondi power. It should be
noted that if Bondi accretion {\it is} responsible for powering radio
jets, so that radio luminosity is correlated with Bondi power, then
objects taken from a flux-limited radio sample are likely to be biased
in the sense that their black hole masses will be above the expected
value for galaxies of their observed properties, given the strong
dependence of $P_{\rm b}$, and thus $Q$, on $M_{\rm BH}$ (eq.\
\ref{mdot}). We make no attempt to correct for this, but it should be
borne in mind in what follows.

For 3C\,31 Laing \& Bridle (2002b) quote a jet power $Q \approx
10^{37}$ W. From the K-band 2MASS observations (here and subsequently
we use the total K-band magnitude provided by the NASA Extragalactic
Database, NED) and the $M_{\rm BH}$ - $K$ relation we infer a black
hole mass of $1.1 \times 10^9$ M$_\odot$. Hardcastle \etal\ (2002,
hereafter H02) studied the small-scale X-ray halo of the host galaxy
of 3C\,31 (the dominant galaxy of a rich group) and measured a central
gas temperature of 0.66 keV. If we assume an efficiency of conversion
of accretion power to jet power of $\eta = 0.1$ [following Allen
\etal: Nemmen \etal\ (2006) have recently argued that efficiencies of
this order are possible in various jet-formation models provided that
the central black hole is rapidly spinning] then we can infer that an
electron density at the Bondi accretion radius (assuming $\rho =
1.13n_{\rm e}m_{\rm p}$ as Allen \etal\ do) $n_{\rm e,A} \approx 6
\times 10^5$ m$^{-3}$ is required to power the jet. H02 give a central
density from their $\beta$-model fit of $n_{\rm e} \approx 2 \times
10^5$ m$^{-3}$, which is already consistent within the uncertainties
on $M_{\rm BH}$: however, we know from the H02 fits to the inner 1.5
arcsec (0.5 kpc) of the 3C\,31 X-ray emission that the density is
higher in that region than the $\beta$-model would predict, and in
fact the mean density inferred from spectral fitting is around $5
\times 10^5$ m$^{-5}$, while clearly if there is any density gradient
in this inner region of the source the density at the Bondi accretion
radius ($r_{\rm A} = 50$ pc) will be higher than the mean value. Thus
there is no difficulty in producing the jets of 3C\,31 by accretion of
the hot phase of the IGM. We have repeated this calculation for
several other FRI radio galaxies for which both X-ray estimates of the
central density (e.g. 3C\,296, Hardcastle \etal\ 2005) and jet power
estimates from jet modelling (Laing, private communication) are
available, and find that this conclusion is generally true: the
central densities in the galaxy-scale components of these sources,
even those in considerably poorer large-scale environments than
3C\,31, are sufficient for Bondi accretion at a nominal 10 per cent
efficiency to power the jet.

\begin{figure*}
\epsfxsize 17cm
\epsfbox{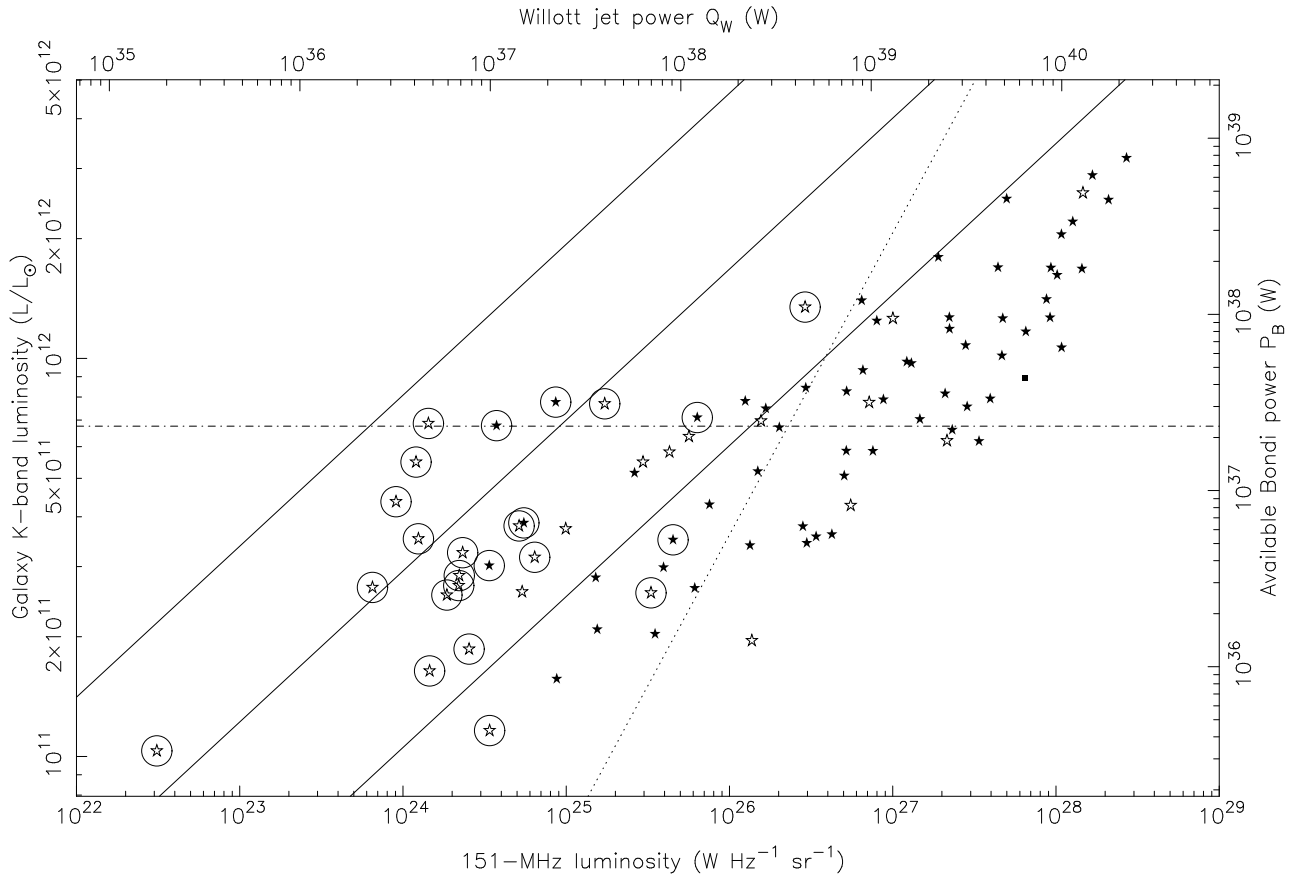}
\caption{K-band host galaxy luminosity against 151-MHz luminosity for
  3CRR narrow-line and low-excitation radio galaxies with available
  K-band magnitudes. The top and right-hand axes show the conversion
  of these observational quantities into jet power derived from the
  Willott \etal\ (1999) relation ($Q_{\rm W}$) and available Bondi
  power ($P_{\rm B}$) respectively, derived from eqs \ref{mdot} and
  \ref{willott} using a single density at the Bondi accretion radius,
  ($\rho_{\rm A} = 5 \times 10^5$ m$^{-3}$), a single temperature at
  that radius ($kT = 0.7$ keV), a single factor $f$ in the Willott et
  al relation ($f=10$) and a single Bondi efficiency ($\eta = 0.1$).
  Black hole masses are inferred from the K-band magnitudes as
  described in the text. Open stars are low-excitation radio galaxies
  and filled stars are narrow-line objects. A circle round a data
  point indicates an FRI. The filled square marks the position of the
  powerful NLRG Cygnus A, as discussed in the text (note that here the
  adopted $K$-band luminosity is derived from the known black hole
  mass rather than vice versa). The central solid line shows equality
  between the predicted Bondi power and the Willott et al jet power.
  The lines on either side are separated from the solid line by one
  order of magnitude, and so give an idea of the scatter expected on
  the Bondi luminosity from the observed dispersion in the $M_{\rm
  BM}$ - $K$ relation. The horizontal dash-dotted line shows the
  galaxy luminosity (corresponding to a black hole mass) at which the
  Bondi accretion rate $\dot M$ for our chosen gas parameters is equal
  to 0.01 times the Eddington rate (so that $P_{\rm B} = 
  10^{-3} L_{\rm Edd}$), while the dotted line shows the line at which
  the Willott jet power $Q_{\rm W}$ equals 0.015 times the Eddington
  luminosity, both (for clarity) assuming the nominal $M_{\rm BM}$ -
  $K$ relation. See the text for discussion of these lines.}
\label{bondi-plot}
\end{figure*}

We next investigate whether there are sources that {\it cannot} be
powered by Bondi accretion under these conditions. A widely used
estimator of jet power is that derived by Willott \etal\ (1999):
\begin{equation}
Q_W = 3 \times 10^{38} f^{3/2} L_{151}^{6/7}\ {\rm W}
\label{willott}
\end{equation}
where $L_{151}$ is the observed radio luminosity in units of $10^{28}$
W Hz$^{-1}$ sr$^{-1}$. The factor $f$ parametrizes our ignorance of
true jet powers, and is likely to depend in practice on the type of
source and its environment, as discussed by Willott \etal . However,
Blundell \& Rawlings (2000) estimate that $f$ may be $\sim 10$ for
FRII sources, while we find by normalizing this relation to jet powers
of FRIs (Laing, private communication) that $f$ lies in the range
10--20 for these objects as well. If we adopt a common $f$ value for
all sources then an object's jet power $Q$ and available Bondi power
$P_{\rm B}$ {\it for given density and temperature at the Bondi
radius} can be calculated simply from its redshift, radio flux density
and $K$-band apparent magnitude (using the $M_{\rm BH}$ - $K$ relation
as discussed above). In Fig.\ \ref{bondi-plot} we plot the
observational quantities, radio luminosity and $K$-band luminosity,
for radio galaxies from 3CRR (Laing, Riley \& Longair 1983) with
available K-band magnitudes, which are taken from the
compilation\footnote{Data available online at\\
http://www-astro.physics.ox.ac.uk/$\sim$cjw/kz/3ckz.html} of Willott
\etal\ (2003) for sources with $z>0.05$ and from 2MASS for nearby
objects. The figure also shows the conversion of the observational
quantities to Bondi power and jet power, assuming values of the
central density and sound speed appropriate for nearby FRI sources.
Sources are divided according to their emission-line classifications
in the manner discussed by H06. Quasars and BLRG are not plotted, as
they have a substantial AGN-related contribution to the observed
K-band magnitude, but we would expect them to behave similarly in
unified models. For reference, we also plot the position of the
high-excitation, powerful non-3CRR object Cygnus A (3C\,405), using
the black hole mass determined by Tadhunter \etal\ (2003).

While of course we do not claim that this plot gives the accurate
position of any given source on the $Q$ - $P_{\rm B}$ plane -- it is
essentially just a version of the well-known magnitude/radio-power
plot of Ledlow \& Owen (1996) -- it is instructive in several ways.
Firstly, it shows that the nearby FRI radio galaxies (for which the
adopted densities and temperatures are comparable to
those directly measured where X-ray observations exist) almost all lie
within a factor of a few of the line of $Q_W = P_{\rm B}$, as we would
expect for the more detailed analysis carried out above. Secondly, it
shows that the majority of low-excitation FRII radio galaxies in our
sample also lie close to this line, within the uncertainties due to
the scatter in the $M_{\rm BH}$ -- $K$ relation. And thirdly, it shows
that there is a population of FRII sources, encompassing most of the
narrow-line FRII sources (and therefore, presumably, all
high-excitation sources), that have jet powers exceeding the available
Bondi powers (for our choice of central gas properties) in some cases
by more than two orders of magnitude. For these sources to be powered
by accretion of the hot IGM they would have to have a much higher
central density than nearby FRIs, $f$ factors much lower than the
values appropriate for the twin-jet FRIs, or central black holes an
order of magnitude more massive than the local $M_{\rm BH}$-$K$
relation predicts. While we cannot rule out high central gas densities
for distant objects, nearby FRII sources in group environments with
detailed {\it Chandra} observations show that the luminosities and
masses of any central hot-gas component in these objects are typically
much {\it lower} than in those of nearby FRI sources (Croston \etal,
in prep.). Thus, for these sources at least, we feel confident in the
claim that Bondi accretion of hot gas almost certainly cannot be
responsible for powering the jet.

We note that several powerful low-excitation radio galaxies
(with estimated jet powers $Q \sim 10^{39}$ W) lie more than an order
of magnitude away from the line of $P_{\rm B} = Q_W$ on Fig.\
\ref{bondi-plot}. One or two of these may not be true low-excitation
objects (H06). Setting that aside, though, we already know that these
powerful FRII LERGs tend to trace cluster environments, much richer
than those of NLRGs, BLRGs and quasars of comparable power
(Hardcastle 2004), and so it is possible that for these objects our
choice of central density (comparable to those in group-centre FRIs)
is too low, which would move them up on Fig.\ \ref{bondi-plot}: at the same
time, their denser environment would tend to lead to higher radio
luminosity for a given jet power (Barthel \& Arnaud 1996), moving them
to the left. Both these effects would bring them closer to the region
of parameter space where accretion from the hot phase could power the
jets. While this picture needs to be tested by detailed X-ray studies
of the environments of powerful LERGs, we do not feel that these
objects present an insuperable problem for the model at present.

Finally, it is worth commenting on one observational point that could
be used to argue against the picture we present here -- the widespread
detection of nuclear dust features in the host galaxies of nearby FRI
sources (e.g. Martel \etal\ 1999). In some cases these structures have
even been described as being related to the accretion disc itself
(e.g.\ Jaffe \etal\ 1993), and, although they are clearly on scales
much too large to be directly related either to any true accretion
disc or to the larger-scale torus, it is possible that they represent
a reservoir of cold gas accreting on to the central black hole.
However, as we know the central cooling rates of the hot gas in these
objects are high (e.g.\ H02) and the jet in classical twin-jet systems
at least is likely to be a relatively inefficient source of heating
for this central hot gas component, it is possible that small-scale
cold material will naturally appear as a result of cooling, mirroring
the processes seen on a larger scale in massive central cluster
galaxies. Tan \& Blackman (2005) argue that formation of a cold disc
with a scale comparable to the Bondi radius is in fact a natural
consequence of Bondi accretion in massive ellipticals. Given these
possibilities, we do not consider that the observations are
inconsistent with the picture presented here.

\section{Implications of the model}
\label{implications}

The analysis of the previous section has shown that it is possible
that all the low-power low-excitation radio galaxies are powered by
accretion from the hot phase, consistent with the fact that their
nuclear spectra show no evidence for cold material close to the
nucleus (i.e., no evidence for the `torus'). On the other
hand, we have seen that narrow-line radio galaxies (and therefore also
broad-line radio galaxies and radio-loud quasars) which have clear
evidence for accretion discs and tori, cannot be powered in this way
--- the large amounts of cold material in the nucleus and the
radiative efficiency of accretion are naturally explained if these
objects are powered by accretion of cold material via a thin disc in
the standard manner.

Does accretion from the hot phase necessarily imply a radiatively
inefficient accretion flow? We know that, by definition (eq.\
\ref{rbondi}) the sound speed of the gas exceeds the Keplerian
velocity $v_k$ at the Bondi radius. For $\gamma = 5/3$, this condition
is maintained throughout the Bondi flow (Bondi 1952). Thus true
spherical Bondi accretion is incompatible with the formation of a thin
disc, which requires $c_s \ll v_k$. However, the effects of a
two-temperature plasma and of viscous dissipation must be considered,
the assumption of spherical symmetry must break down at some point,
and adequate radiatively inefficient cooling models close to the black
hole should probably resemble a quasi-spherical accretion-dominated
advection flow (ADAF: Narayan \& Yi 1995a,b) or one of the numerous
variants discussed in the literature. We know that these solutions in
general are inconsistent with accretion rates of order of or greater
than the Eddington rate, which we define as $L_{\rm Edd}/c^2$, i.e.
\begin{equation}
\dot M_{\rm Edd} = {{4\pi GM_{\rm BH} m_p}\over{\sigma_{\rm T} c}}
\label{eddington}
\end{equation}
where $\sigma_{\rm T}$ is the Thomson cross-section and $m_p$ is the
mass of a proton. Therefore an important check on whether our picture
is consistent with ADAF-type solutions is to ask whether it requires
accretion rates of the order of the Eddington rate. Comparing eqs
\ref{mdot} and \ref{eddington} it can be seen that $\dot m = \dot
M_{\rm B}/\dot M_{\rm Edd}$ is linear in black hole mass for given
parameters of the external gas, and that the black hole mass required
to give $\dot m \sim 1$ is high for plausible external thermal
parameters. In fact, all of the black hole masses for the
low-excitation sources give $\dot m \ll 1$ for our choice of Bondi
parameters. The maximum value of $\dot m$ for a low-excitation source
close to the line of $P_{\rm B} = Q_{\rm W}$ in Fig.\ \ref{bondi-plot}
is $\sim 0.02$, and most lie below $\dot m = 0.01$. To illustrate this
we have plotted the black hole mass (and therefore galaxy mass)
corresponding to $\dot m = 0.01$ for our hot-gas parameters as the
dash-dotted line in Fig.\ \ref{bondi-plot}. Thus all the
low-excitation objects consistent with being powered by Bondi
accretion (i.e. within the solid lines in Fig.\ \ref{bondi-plot}) also
have significantly sub-Eddington accretion rates, as required by
ADAF-type solutions. In addition, we have plotted (dotted line) the
line of $Q_{\rm W} = 0.015L_{\rm Edd}$ for the inferred black hole
masses. This line of jet power vs. Eddington luminosity was used to
divide FRI and FRII sources by Ghisellini \& Celotti (2001). In the
present plot we would argue that, while it does indeed separate FRIs
and FRIIs reasonably well, it (or any variant on it represented by
shifting the line to right or left) separates the low-excitation and
high-excitation objects less well than the Bondi lines: in particular,
there is a class of high-excitation low-power FRIIs with low $Q_{\rm
W}/L_{\rm Edd}$ that could not lie to the right of such a line without
also including a number of low-excitation FRIIs. Although the sample
size is very small, this provides some weak evidence that it is
genuinely the origin of the accreting material, and not simply the
value of $Q_{\rm W}/L_{\rm Edd}$, that determines the accretion mode
of radio galaxies (cf.\ section 5.2 of Narayan \& Yi 1995b).

In the rest of this section of the paper we therefore explore some of
the consequences of a picture in which there is a causal connection
between the origin of the accreting gas and the accretion mode, as
indicated by the emission-line type of the galaxy. We know that most,
though not all, low-power radio galaxies are low-excitation objects,
while essentially all the high-power radio galaxies (and of course all
powerful radio-loud quasars) are high-excitation objects. Thus it is
not too much of a simplification to say that low-power radio galaxies
(roughly in the FRI regime, but including a handful of less powerful
FRIIs) are likely to be powered by `hot-mode' accretion, while
powerful FRII radio galaxies and quasars in general will trace
`cold-mode' accretion. This allows us to interpret some previous
results that have been stated in terms of luminosity differences, or
FRI/FRII differences, as being more naturally understood in terms of a
dichotomy in accretion mode, bearing in mind that the FRI/FRII
difference, in our picture, is a result of the jets' interaction with
the large-scale environment rather than a direct consequence of the
nature of the accretion.

\subsection{Feedback}

Feedback from AGN outbursts is now thought to be an important
ingredient in galaxy formation models (e.g. Croton \etal\ 2006),
enabling successful reproduction of the high-mass end of the galaxy
luminosity function and solving the `cooling flow' problem in the
centre of massive clusters. An important feature of current models in
which AGN heating prevents cooling of the ICM in the centre of massive
clusters is that the AGN should both be able to influence, {\it and
should be influenced by}, the X-ray emitting phase. Direct accretion
of the hot phase provides an elegant way of ensuring that the AGN
activity is regulated by the gas properties at the cluster centre,
which was of course the motivation for the work of Allen \etal\
(2006). It is clear, though, that this is only possible for a
`hot-mode' radio source. Radio galaxies and quasars accreting in the
cold mode do not have this direct connection between the hot phase and
the rate of fuelling of the AGN: instead, the jet power is controlled
solely by the accretion rate of cold gas, which may have nothing to do
with the state of the hot phase. It is thus possible for cold-mode
sources to inject catastrophic amounts of energy into the hot phase of
the IGM. This is borne out by recent studies of the poor environments
of nearby NLRG FRIIs (Kraft \etal\ 2007; Hardcastle \etal\ 2007) that
show that the work done by the radio sources plus the internal energy
in their lobes of the radio sources is comparable to the entire
thermal energy of their host poor-group environments. Cold-mode
systems can nevertheless play some role in feedback models. Churazov
\etal\ (2005) suggest an evolutionary feedback model in which
elliptical galaxies go through an early stage of rapid black hole
growth at high accretion rates, until the black hole becomes
sufficiently massive that radio-mode feedback turns on, slowing the
rate of black hole growth. The radio outbursts of cold-mode systems
may occur at an intermediate stage in this evolution, rather than
forming part of an eventual feedback loop; the un-self-regulated
energy input from this type of outburst into the ICM may instead be
responsible for the ``entropy excess'' observed in galaxy groups and
clusters (e.g. Pratt, Arnaud \& Pointecouteau 2006).

\subsection{Environments of active galaxies}

In the model we have outlined we expect different types of
active galaxies to be found in different environments. Cold-mode
accretion requires a supply of cold gas: the easiest way for an
elliptical galaxy to acquire this is by a merger with a gas-rich system.
Samples of high-excitation radio galaxies should thus
show evidence for mergers and interactions, consistent with many
observations showing evidence for recent or ongoing mergers in the
hosts of powerful sources (e.g. Heckman \etal\ 1986). Of course, since
the timescale for transport of cold gas to the galactic centre may be
much longer than the timescale for removal of the obvious optical
signature of a merger, we do not expect a one-to-one correlation
between observed elliptical-spiral mergers and AGN activity. We note,
though, that this picture is consistent with observations of the few
low-power, FRI, radio galaxies that we know to have heavily obscured
nuclear X-ray emission, including Cen A (e.g. Evans \etal\ 2004) and
NGC 3801 (Croston, Kraft \& Hardcastle 2007). Host galaxies of
cold-mode systems do not need a rich environment, or to be at the
bottom of a deep potential well, so long as galaxy-galaxy mergers can
take place.

By contrast, hot-mode accretion requires a supply of hot gas and a
massive central black hole. Both the black hole mass and the mass of
the galaxy-scale X-ray halo (e.g. Mathews \& Brighenti 2003) are
correlated with the mass of the host galaxy. Thus we expect hot-mode
systems -- which, observationally, include almost all FRI radio
galaxies -- to favour massive galaxies, and the most powerful radio
sources to tend to be group- or cluster-dominant systems, as is
observed (e.g. Longair \& Seldner 1979; Prestage \& Peacock 1988; Owen
\& White 1991). In samples that are not radio-selected, and in which
the radio population (given the luminosity function) will therefore be
dominated by low-luminosity radio galaxies accreting in the hot mode,
we expect a strong correlation between radio activity and host
galaxy/black hole mass, as found by Best \etal\ (2005). Best \etal\
(2006) have already pointed out the potential link between this
observation and the fuelling of these radio sources by accretion of
the hot phase, but in our picture their findings cannot be generalized
to high-power, high-excitation, cold-mode radio sources.

\subsection{AGN populations}

The luminosity function of hot-mode radio galaxies will be determined
by a combination of the black hole mass function and the distribution
of properties of central hot gas. The cold-mode luminosity function,
on the other hand, will be determined by the rate of accretion of cold
material on to the central black hole, with no direct effect of the
black hole mass until the Eddington luminosity is reached. We
therefore expect the luminosity functions to be different, and the
local radio luminosity function of radio-loud AGN, being the composite
of two different luminosity functions, should have a break at a
luminosity comparable to the transition between populations dominated
by low-excitation and high-excitation radio sources, as is observed
(e.g. Machalski \& Godlowski 2000). The `dual-population' unified
model of Jackson \& Wall (1999) proposes that the luminosity functions
of FRI and FRII populations evolve differently with cosmic time. Since
we know that the conditions for hot-mode and cold-mode accretion {\it
must} vary with time, we would expect that this type of model would
more properly be applied to the low-excitation and high-excitation
sources.

\subsection{Jet production and accretion history}

We have so far not attempted to comment on the formation of jets in
these systems. In our picture, the FRI/FRII difference is not in
origin a function of accretion mode: the FRII LERG
population provides the clearest evidence for this feature of the
model. While the nuclear and large-scale properties are correlated in
general (in 3C\,31 and other twin-jet FRI sources, for example, the
dense central concentration of hot gas required to fuel the active
nucleus in the hot mode also provides the pressure gradient needed to
collimate the jet and keep it stable to large distances: H02) they
need not be in particular cases, which explains the lack of a
one-to-one relationship between accretion mode and FR class. It is
thus a requirement of the model that the two accretion modes must be
capable of producing jets that are similar in most of their observable
parsec-scale properties, as FRI and FRII jets are known to be (e.g.
Pearson 1996). In models where the jet is produced from the accretion
flow (Blandford \& Payne 1982) it seems likely that this would require
similar structure in the innermost regions of the accretion flow, but
clearly we cannot distinguish between particular jet formation models.

A potentially more interesting question is whether hot-mode sources
{\it necessarily} form a jet. Cold-mode sources can be divided into
radio-loud and radio-quiet classes: is the same true for hot-mode
accretion? At least some hot-mode sources at cluster centres require a
more or less continuously active jet, to reproduce the nearly
universal detection of radio galaxies in high cooling-rate cluster
cores (e.g. Eilek \& Owen 2006) and it may be this is true for
hot-mode sources in general: that is, it may be that this mode of
accretion always produces a jet. Certainly it has been argued (e.g.\
Ho 2002; Nagar \etal\ 2002) that low-power AGN (with low values of
$L_{\rm bol}/L_{\rm Edd}$), where identified as AGN, tend to be
dominated by jet emission. However, it would be hard to identify a
class of AGN accreting in the hot mode in a radio-quiet manner,
without powerful jets -- we only know about the active nuclei in LERGs
because of their jet-related radio, optical and X-ray emission. The
only accretion-related radiation in this situation would be the
extremely weak emission from the radiatively inefficient flow itself,
but these sources would still contribute to black hole growth.
Although it is generally argued that radiatively inefficient accretion
has only a small effect on the evolution of the black hole mass
function (BHMF), direct estimates of its effect (e.g. Merloni \etal\
2004; Hopkins, Narayan \& Hernquist 2006) are based on samples of AGN
that are detectable as such, either by radio or optical emission, and
so might not include jetless hot-mode objects. Sample-independent
constraints on the effects of radiatively inefficient accretion
independent of the population of objects being considered come from
work like that of Cao (2007), who shows that radiatively inefficient
accretion cannot be very important in the evolution of the BHMF if it
is not to overproduce the hard X-ray background, but this relies on
theoretical assumptions about the spectral energy distribution of a
radiatively inefficient accretion flow, possibly still leaving some
loophole for hot-mode objects. In any case, we would expect that if
radiatively inefficient accretion does have any effect on the BHMF, it
will do so primarily in sources where the Bondi accretion rate is
high. This would predict an environmental dependence of the BHMF,
though the effect is probably too weak to detect at present.

\subsection{Analogy with X-ray binaries}

Recently there has been considerable discussion of the relationship
between the accretion modes and jet behaviour of X-ray binaries
(`microquasars'; hereafter XRB) and those of AGN (e.g.\ Fender,
Belloni \& Gallo 2004, K\"ording, Jester \& Fender 2006). Support for
a connection between XRB jet/accretion states and those of various
different types of AGN comes from the so-called `fundamental plane of
black-hole activity' (e.g. Merloni, Heinz \& Di Matteo 2003, Falcke,
K\"ording \& Markoff 2004), and from similarities in their variability
properties (e.g.\ McHardy \etal\ 2006). Radio-quiet AGN are usually
associated with the high $\dot M$ high/soft spectral state of XRB, and
low-power (FRI) radio galaxies with the low/hard state (e.g.
Maccarone, Gallo \& Fender 2003). Powerful FRII radio galaxies and
radio-loud quasars have sometimes been associated with the radio
outbursts seen as XRB transition between states (e.g. K\"ording,
Jester \& Fender 2006). Churazov et al. (2005) recently suggested that
the central AGN of massive elliptical galaxies evolve through an early
stage of radiatively efficient, high accretion rate black hole growth
(analagous to the radio-quiet high/soft XRB state) to a state of
stability regulated by feedback from radiatively inefficient feedback
(analogous to the jet-dominated low/hard state).

In the picture of radio source activity we have presented, the two
different modes of accretion (radiatively inefficient accretion of the
hot medium and radiatively efficient accretion of cold gas) {\it both}
have to operate to produce radio jets; in this model the jet
properties are essentially independent of the accretion mode. There is
a strong relationship between radio morphology and accretion mode in
our picture, but this is likely to be due to the need for an accretion
rate higher than that available from Bondi accretion of the hot phase
(in most environments) in order to produce powerful jets. By removing
the direct connection between accretion mode and radio-jet properties,
our model causes some difficulties for a simple connection between XRB
and AGN jet/disc states. In our picture, the low-excitation radio
galaxies (including the FRIs) can be straightforwardly associated with
the radiatively inefficient low/hard XRB state; however, it is unclear
how the high-excitation sources fit into the picture, since they are
clearly undergoing radiatively efficient accretion with steady jet
emission. It remains possible that high-excitation sources are in some
sense transitional objects (as discussed above, many such systems are
associated with mergers, which plausibly cause a dramatic increase in
accretion rate), but these systems can maintain steady jets for $\sim
10^{7}$ years in a radiatively efficient accretion mode, so that it is
not clear whether a direct analogy with XRB behaviour can really be
made for the high-power, cold-mode systems.

\section{Summary and future work}
\label{summary}

The idea that low-excitation radio galaxies are fuelled by the
accretion of the hot, X-ray emitting phase of the IGM, while
high-excitation radio sources are powered by accretion of cold
material, can be used (qualitatively) to explain

\begin{itemize}
\item their different optical and X-ray nuclear properties (H06, this
  paper)
\item the close relationship between the power output of low-power radio
  galaxies and the energy needed to solve the cooling flow problem
  (Allen \etal\ 2006)
\item the association of low-power radio galaxies with the most
  massive host galaxies (Best \etal\ 2006)
\item the observed differences in cosmic evolution of low- and
  high-power radio-loud AGN (this paper), and
\item the different environments of low- and high-excitation radio
  sources (this paper)
\end{itemize}

In this paper we have shown quantitatively that the required
difference between low-excitation and high-excitation objects in the
well-studied 3CRR sample is at least plausible. Clearly, though,
further testing of this picture is essential. Observational tests
would include a significant expansion of the available database of
sensitive, high spatial resolution X-ray spectroscopic observations of
nuclei of radio galaxies -- our conclusions on the nature of
low-excitation FRII sources in particular are based on small samples.
Better constraints on the small-scale hot-gas environments of radio
sources are vital, and forthcoming mid-infrared studies of the nuclei
of radio-loud AGN will also provide important information. In future
we expect studies of the evolution of the luminosity functions of the
two populations, combined with cosmological simulations that give
indications of the availability of fuel to the two accretion
processes, to provide the most stringent tests of the model.

\section*{acknowledgements}

We thank Robert Laing for communicating estimated jet powers for FRI
sources in advance of publication, Marc Sarzi for information on black
hole mass estimates, Danny Steeghs and Jeff McClintock for helpful
discussion of X-ray binaries, and Steve Allen, Philip Best, Sebastian
Jester, Christian Kaiser and Elmar K\"ording for discussion of the
ideas presented in this paper at various times over the past year. We
also thank the anonymous referee for insightful comments that helped
us to improve the paper. MJH thanks the Royal Society for a Research
Fellowship. This research has made use of the NASA/IPAC Extragalactic
Database (NED) which is operated by the Jet Propulsion Laboratory,
California Institute of Technology, under contract with the National
Aeronautics and Space Administration.


\begin{thebibliography}{}
\bibitem[]{17}Allen, S.W., Dunn, R.J.H., Fabian, A.C., Taylor, G.B., Reynolds, C.S., 2006, MNRAS, 372, 21
\bibitem[]{55}Barthel, P.D., 1989, ApJ, 336, 606
\bibitem[]{57}Barthel, P.D., Arnaud, K.A., 1996, MNRAS, 283, L45
\bibitem[]{65}Baum, S.A., Zirbel, E.L., O'Dea, C.P., 1995, ApJ, 451, 88
\bibitem[]{67}Belsole, E., Worrall, D.M., Hardcastle, M.J., 2006, MNRAS, 336, 339
\bibitem[]{79}Best, P.N., Kauffmann, G., Heckman, T.M., Brinchmann, J., Charlot, S., Ivezi\'c, Z., White, S.D.M., 2005, MNRAS, 362, 25
\bibitem[]{80}Best, P.N., Kaiser, C.R., Heckman, T.M., Kauffmann, G., 2006, MNRAS, 368, L67
\bibitem[]{112}Blandford, R.D., Payne, D.G., 1982, MNRAS, 199, 883
\bibitem[]{120}Blundell, K.M., Rawlings, S., 2000, AJ, 119, 1111 
\bibitem[]{129}Bondi, H., 1952, MNRAS, 112, 195
\bibitem[]{196}Canvin, J.R., Laing, R.A., Bridle, A.H., Cotton, W.D., 2005, MNRAS, 363, 1223
\bibitem[]{197}Cao, X., 2006, ApJ ~in press (astro-ph/0701007)
\bibitem[]{221}Chiaberge, M., Capetti, A., Celotti, A., 2002, A\&A, 394, 791
\bibitem[]{225}Churazov, E., Sazonov, S., Sunyaev, R., Forman, W., Jones, C., B\"ohringer, H., 2005, MNRAS, 363, L91
\bibitem[]{276}Croston, J.H., Kraft, R.P., Hardcastle, M.J., 2007, ApJ submitted
\bibitem[]{277}Croton, D., et al., 2006, MNRAS, 365, 111
\bibitem[]{322}Eilek, J.A., Owen, F.N., 2006, in B\"ohringer H., Schuecker P., Pratt G.W. \& Finoguenov A., eds, Heating vs. cooling in galaxies and clusters of galaxies, Springer-Verlag, Heidelberg, astro-ph/0612111
\bibitem[]{333}Evans, D.A., Kraft, R.P., Worrall, D.M., Hardcastle, M.J., Jones, C., Forman, W.R., Murray, S.S., 2004, ApJ, 612, 786
\bibitem[]{334}Evans, D.A., Worrall, D.M., Hardcastle, M.J., Kraft, R.P., Birkinshaw, M., 2006, ApJ, 642, 96
\bibitem[]{344}Falcke, H., K\"ording, E., Markoff, S., 2004, A\&A, 414, 895
\bibitem[]{362}Fender, R.P., Belloni, T.M., Gallo, E., 2004, MNRAS, 355, 1105
\bibitem[]{401}Ghisellini, G., Celotti, A., 2001, A\&A, 379, L1
\bibitem[]{439}Haas, M., et al., 2004, A\&A, 424, 531
\bibitem[]{450}Hardcastle, M.J., 2004, A\&A, 414, 927
\bibitem[]{461}Hardcastle, M.J., Evans, D.A., Croston, J.H., 2006, MNRAS, 370, 1893
\bibitem[]{462}Hardcastle, M.J., Worrall, D.M., Birkinshaw, M., Laing, R.A., Bridle, A.H., 2002, MNRAS, 334, 182 [H02]
\bibitem[]{464}Hardcastle, M.J., Kraft, R.P., Worrall, D.M., Croston, J.H., Evans, D.A., Birkinshaw, M., Murray, S.S., 2007, ApJ  submitted
\bibitem[]{470}Hardcastle, M.J., Worrall, D.M., 1999, MNRAS, 309, 969
\bibitem[]{475}Hardcastle, M.J., Worrall, D.M., Birkinshaw, M., Laing, R.A., Bridle, A.H., 2005, MNRAS, 358, 843
\bibitem[]{498}Heckman, T.M., Smith, E.P., Baum, S.A., van Breugel, W.J.M., Miley, G.K., Illingworth, G.D., Bothun, G.D., Balick, B., 1986, ApJ, 311, 526
\bibitem[]{512}Hine, R.G., Longair, M.S., 1979, MNRAS, 188, 111
\bibitem[]{521}Ho, L.C., 2002, ApJ, 564, 120
\bibitem[]{522}Hopkins, P.F., Narayan, R., Hernquist, O.L., 2006, ApJ, 643, 641
\bibitem[]{546}Jackson, C.A., Wall, J.V., 1999, MNRAS, 304, 160
\bibitem[]{551}Jackson, N., Rawlings, S., 1997, MNRAS, 286, 241
\bibitem[]{554}Jaffe, W., Ford, H.C., Ferrarese, L., van de Bosch, F., O'Connell, R.W., 1993, Nat, 364, 213
\bibitem[]{601}K\"ording, E.G., Jester, S., Fender, R., 2006, MNRAS, 372, 1366
\bibitem[]{622}Kraft, R.P., Birkinshaw, M., Hardcastle, M.J., Evans, D.A., Croston, J.H., Worrall, D.M., Murray, S.S., 2007, ApJ ~in press (astro-ph/070701458)
\bibitem[]{642}Laing, R.A., Bridle, A.H., 2002a, MNRAS, 336, 328
\bibitem[]{643}Laing, R.A., Bridle, A.H., 2002b, MNRAS, 336, 1161
\bibitem[]{645}Laing, R.A., Canvin, J.R., Bridle, A.H., Hardcastle, M.J., 2006, MNRAS, 372, 510
\bibitem[]{646}Laing, R.A., Jenkins, C.R., Wall, J.V., Unger, S.W., 1994, in Bicknell G.V., Dopita M.A., Quinn P.J., eds, The First Stromlo Symposium: the Physics of Active Galaxies, ASP Conference Series vol. 54, San Francisco, p.~201
\bibitem[]{651}Laing, R.A., Riley, J.M., Longair, M.S., 1983, MNRAS, 204, 151
\bibitem[]{681}Ledlow, M.J., Owen, F.N., 1996, AJ, 112, 9
\bibitem[]{712}Longair, M.S., Seldner, M., 1979, MNRAS, 189, 433
\bibitem[]{728}Maccarone, T.J., Gallo, E., Fender, R., 2003, MNRAS, 345, L19
\bibitem[]{733}Machalski, J., Godlowski, W., 2000, A\&A, 360, 463
\bibitem[]{750}Marconi, A., Hunt, L.K., 2003, ApJ, 589, L21
\bibitem[]{757}Martel, A.R., et al., 1999, ApJS, 122, 81
\bibitem[]{766}Mathews, W.G., Brighenti, F., 2003, ARA\&A, 41, 191
\bibitem[]{775}McHardy, I.M., K\"ording, E., Knigge, C., Uttley, P., Fender, R.P., 2006, Nat, 444, 730
\bibitem[]{779}Merloni, A., 2004, MNRFAS 353 1035
\bibitem[]{780}Merloni, A., Heinz, S., Di, Matteo, T., 2003, MNRAS, 345, 1057
\bibitem[]{813}Nagar, N.M., Falcke, H., Wilson, A.S., Ulvestad, J.S., 2002, A\&A, 392, 53
\bibitem[]{815}Narayan, R., Yi, I., 1995a, ApJ, 444, 231
\bibitem[]{816}Narayan, R., Yi, I., 1995b, ApJ, 452, 710
\bibitem[]{817}Nemmen, R.S., Bower, R.G., Babul, A., Storchi-Bergmann, T., 2006, MNRAS in press, astro-ph/0612354
\bibitem[]{839}Ogle, P., Whysong, D., Antonucci, R., 2006, ApJ in press, astro-ph/0601485
\bibitem[]{863}Owen, F.N., White, R.A., 1991, MNRAS, 249, 164
\bibitem[]{883}Pearson, T.J., 1996, in Hardee P.E., Bridle A.H., Zensus J.A., eds, Energy Transport in Radio Galaxies and Quasars, ASP Conference Series vol.~100, San Francisco, p.~97
\bibitem[]{916}Pratt, G.W., Arnaud, M., Pointecouteau, E., 2006, A\&A, 446, 429
\bibitem[]{918}Prestage, R.M., Peacock, J.A., 1988, MNRAS, 230, 131
\bibitem[]{1094}Tadhunter, C., Marconi, A., Axon, D., Wills, K., Robinson, T.G., Jackson, N., 2003, MNRAS, 342, 861
\bibitem[]{1096}Tan, J.C., Blackman, E.G., 2005, MNRAS, 362, 983
\bibitem[]{1120}Tremaine, S., et al., 2002, ApJ, 574, 740
\bibitem[]{1170}Whysong, D., Antonucci, R., 2004, ApJ, 602, 116
\bibitem[]{1183}Willott, C.J., Rawlings, S., Blundell, K.M., Lacy, M., 1999, MNRAS, 309, 1017
\bibitem[]{1184}Willott, C.J., Rawlings, S., Jarvis, M.J., Blundell, K., 2003, MNRAS, 339, 173
\end{thebibliography}
\end{document}